\documentclass[pra,showkeys,showpacs,groupedaddress,twocolumn]{revtex4-1}
\usepackage{amssymb,comment}
\usepackage{epsfig,amsmath,graphicx}
\usepackage[usenames]{color}
\usepackage[colorlinks,linkcolor=blue,anchorcolor=blue,citecolor=blue]{hyperref}
\usepackage{txfonts}
\usepackage{bm}

\newcommand{\be}[0]{\begin{equation}}
\newcommand{\ee}[0]{\end{equation}}

\def\bea{\begin{eqnarray}}
\def\eea{\end{eqnarray}}
\def\bes{\begin{subequations}}
\def\ees{\end{subequations}}

\allowdisplaybreaks[4]

\begin{document}

\title{Stable 3D vortex solitons of high topological charge in a Rydberg-dressed Bose-Einstein condensate with spin-orbit coupling}
\author{Yanchao Zhang$^{1}$}
\author{Chao Hang$^{1,2,3}$}\thanks{chang@phy.ecnu.edu.cn}
\author{Boris A. Malomed$^{4,5}$}
\author{Guoxiang Huang$^{1,2,3}$}\thanks{gxhuang@phy.ecnu.edu.cn}
\affiliation{$^1$State Key Laboratory of Precision Spectroscopy, East China Normal University, Shanghai 200241, China}
\affiliation{$^2$NYU-ECNU Institute of Physics, New York University at Shanghai, Shanghai 200062, China}
\affiliation{$^3$Collaborative Innovation Center of Extreme Optics, Shanxi University, Taiyuan, Shanxi 030006, China}
\affiliation{$^4$Department of Physical Electronics, School of Electrical Engineering, Faculty of Engineering, Tel Aviv University, Tel Aviv, Israel}
\affiliation{$^5$Instituto de Alta Investigaci\'{o}n, Universidad de Tarapac\'{a}, Casilla 7D, Arica, Chile}
%

\begin{abstract}
Stable vortex solitons  (VSs) are objects of great interest for fundamental studies and various applications, including particle trapping, microscopy, data encoding, and matter-wave gyroscopes. However, three-dimensional (3D) VSs with high topological charges, supported by self-attractive nonlinearities, are unstable against fragmentation, which eventually leads to internal blowup (supercritical
collapse) of the fragments. Here, we propose a scheme for realizing stable 3D VSs with topological charges up to $5$ and $6$ in the two components of a binary, Rydberg-dressed Bose-Einstein condensate (BEC) with spin-orbit coupling (SOC). We show that, if the SOC strength exceeds a critical value, the rotational symmetry of the VSs in the transverse plane gets broken, resulting in separation of the two components. Nevertheless, the VSs with the broken symmetry remain stable. The VS stability domains are identified in the system's parameter space. Moreover, application of torque to the stable VSs sets them in the state of robust gyroscopic
precession.
\end{abstract}

\maketitle

\textit{Introduction.-} Vortex solitons (VSs), i.e. self-trapped
localized modes of nonlinear fields with embedded vorticity,
have attracted a lot of interest in the course of the past few decades~\cite{Pismen1999}. Comparing with fundamental solitons, the intrinsic vorticity carried by VSs is characterized by an integer topological charge (winding number) $S$,
which is defined through a total change of the phase equal to $2\pi S$
produced by a round trip along a closed trajectory surrounding the vortex
pivot (phase singularity). In addition to their significance for fundamental
studies, VSs are promising for applications, such as particle
trapping, data encoding, microscopy, controllable angular momentum transfer from light to matter, etc.~\cite{Mihalache2021,Malomed2019}.

Although VSs were predicted in many physical systems, ranging
from Bose-Einstein condensates (BECs)~\cite{Abo-Shaeer2001,Anderson2001,Keeling2008,Lagoudakis2008,Tikhonenkov2008,
Fetter2009,Sanvitto2010,Gao2018,Berger2020,Sitnik2022,Rosen2022}
to optical systems~\cite{Duree1995,Scheuer1999,Neshev2004,Fleischer2004,
Desyatnikov2005,Piccardo2022,Zhang2022,Zhao2023}, their stability remains a great challenge even for the lowest topological
charge, $S=1$. In addition to the well-known critical and supercritical
collapse, driven by ubiquitous cubic self-focusing respectively in
two-dimensional (2D) and 3D spaces~\cite{Berge,Fibich}, VSs are
still subject to stronger azimuthal instability, which breaks axially
symmetric 2D vortex rings or 3D vortex tori into fragments, each one being a
fundamental soliton~\cite{YVKartashov2019,Malomed2019}. Many schemes were
proposed for stabilizing VSs, e.g., competing~\cite{Teixeiro1997,Teixeiro1999,Desyatnikov2000,Towers2001,Reyna2016} and
nonlocal~\cite{Briedis2005,Yakimenko2005,Rotschild2005,Rotschild2015}
nonlinearities, various external potentials~\cite{Yang2003,Neshev2004,Fleischer2004,Kartashov2005a,Kartashov2005b,
Terhalle2008,Law2009,Dong2011,Pryamikov2021}, and other mechanisms~\cite{YVKartashov2019,Malomed2019}. In addition to
conservative systems, stable VSs can also be supported in dissipative
systems with the help of localized gain~\cite{Mihalache2006,Leblond2009,Soto-Crespo2009,Skarka2010,Kartashov2010,Lobanov2011,
Borovkova2011,Huang2013,Veretenov2016,Veretenov2017,Li2024}.

In contrast with numerous studies on 2D VSs, 3D vortex modes (toroidal states) were explored less frequently, as their stabilization is a more challenging
problem~\cite{Desyatnikov2000,Mihalache2006,YVKartashov2019,Malomed2019}.
Therefore, the elaboration of new stabilization mechanisms for 3D VSs
remains a relevant aim. Relatively recently, the action of spin-orbit
coupling (SOC), in the form of linear mixing between two components of
binary BEC through the first spatial derivatives of their wavefunctions,
has been realized in atomic BECs~\cite{Lin2009,Lin2011,Dalibard2011}. Then,
it was demonstrated theoretically that SOC provides an efficient mechanism
for stabilizing fundamental and vortical solitons in 1D~\cite%
{YXu2013,Achilleos2013a,Achilleos2013b,Kartashov2013}, 2D~\cite%
{Fetter2014,Salasnich2014,Sakaguchi2014a,Xu2015,Jiang2016,dip-dip,Bin,
Kartashov2019,Adhikari2021,Chenzhou}%
, and 3D~\cite{Zhang2015,Gautam2018} settings with spin-1/2 (two-component)
and spin-1 (three-component) BECs.

In this Letter, we propose a scheme for stabilizing high-order
3D VSs in a SO coupled atomic BEC, dressed by a Rydberg state~\cite{Heidemann2008,Henkel2013}, which provides strong long-range Rydberg interaction between atoms~\cite{Gallagher2008,Sibalic2018,Mohapatra2007,Saffman2010}.
We show that pairs of 3D VSs with different pseudo-spins and winding numbers up to
$S_1=5$ and $S_2=6$ in the two components can be stabilized. We also show that,
when the SOC strength exceeds a critical value, the rotational symmetry of the
VSs in the transverse plane will be broken, leading to spatial separation between the
two spin components. The stability diagrams of the VSs with different topological charges are identified in parameter space by stability analysis and numerical simulations. Moreover, we demonstrate that the stable 3D VSs predicted here can realize robust gyroscopic dynamics~\cite{Driben2014a,Driben2014b}, which may be used, in particular, for measuring features of Rydberg states.

\begin{figure}[tbh]
\centering
\includegraphics[width=0.85\columnwidth]{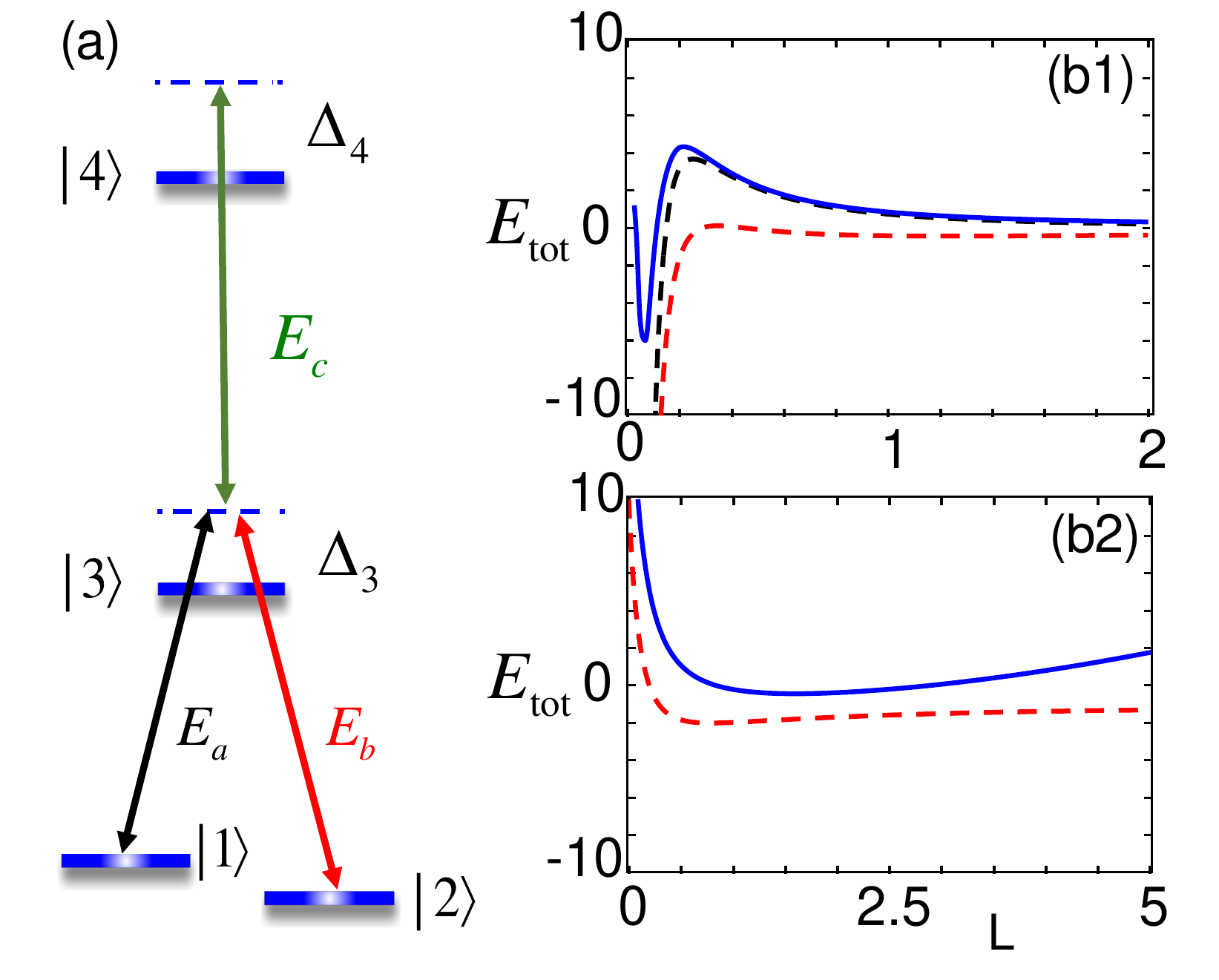}
\caption{(a)~Level diagram and excitation scheme of the Rydberg-dressed BEC.
Ground states $|1\rangle$ and $|2\rangle$ are coupled to excited state $|3\rangle$ by laser fields $\mathbf{E}_{a}$ and $\mathbf{E}_{b}$, respectively. $|3\rangle$ is dressed by Rydberg state $|4\rangle$ by control field $\mathbf{E}_{c}$.
(b1)~$E_{\mathrm{tot}}$ in the weakly nonlocal regime as a function of $L$, obtained
for $c_{\mathrm{kin}}=1$ and $c_{\mathrm{int,0}}=0.2$.
Black dashed, red dash-dotted, and blue solid lines are for $[c_{\mathrm{soc}}k_{L},\,c_{\mathrm{int,1}}]=[0,0]$,
$[c_{\mathrm{soc}}k_{L},\,c_{\mathrm{int,1}}]=[1.2,0]$, and
$[c_{\mathrm{soc}}k_{L},\,c_{\mathrm{int,1}}]=[0,0.01]$,
respectively.
(b2)~$E_{\mathrm{tot}}$ in the strongly nonlocal regime as a function of $L$, obtained for $c_{\mathrm{kin}}=1$.
Red dash-dotted and blue solid lines are for
$[c_{\mathrm{soc}}k_{L},\,c_{\mathrm{int,2}}]=[1.2,0]$ and
$[c_{\mathrm{soc}}k_{L},\,c_{\mathrm{int,2}}]=[0,0.1]$,
respectively.
For more detail, see text.
}
\label{fig1}
\end{figure}

\textit{Model.-} We consider a cold $^{87}$Rb atomic gas, with the atoms Bose-condensed in $F=1$ hyperfine ground state. A static and constant magnetic field $B_0$ is used to split the ground state into three Zeeman sublevels $|5S_{1/2},F=1,m_{F}=-1\rangle$, $|5S_{1/2},F=1,m_{F}=0\rangle $, and $|5S_{1/2},F=1,m_{F}=1\rangle $. The gas is also illuminated by two Raman laser fields $\mathbf{E}_{\alpha}$  (wave vectors $\mathbf{k}_{\alpha }$, angular
frequencies $\omega _{\alpha }$, half-Rabi frequencies $\Omega _{\alpha }$; $\alpha =a,b$), driving transitions from ground-state sublevels $|1\rangle =|5S_{1/2},F=1,m_{F}=-1\rangle $ and $|2\rangle=|5S_{1/2},F=1,m_{F}=0\rangle $ to a common excited level $|3\rangle=|6P_{3/2}\rangle $, respectively [see Fig.~\ref{fig1}(a)].
We assume that the frequency splitting between $|1\rangle $ and $|2\rangle $ is close to the frequency difference between the two Raman lasers, thus the
two-photon detuning $\Delta _{2}=\omega _{b}-\omega _{a}-(\omega _{2}-\omega_{1})\approx 0$.

In such a system, the states $|1\rangle$ and $|2\rangle$ play a role of two pseudo-spin components, which linearly interact through a Rashba-type SOC. The SOC of the Rashba or Dresselhaus type may be synthesized by using a time-dependent gradient magnetic field~\cite{ZFXu2013,Anderson2013} or Raman laser dressing~\cite{Wu2016,Wang2018}; for more details, see Sec.~S1 of
Supplementary material (SM)~\cite{SM}. To get a strong, long-range interaction between
the atoms, the state $|3\rangle $ is assumed to be dressed by a high-lying Rydberg state $|4\rangle =|nD_{3/2}\rangle $ ($n=65$ is principle quantum number) through using a coupling of control laser field $\mathbf{E}_{c}$ (wave vector $\mathbf{k}_{c}$, angular frequency $\omega _{c}$, half-Rabi
frequency $\Omega _{c}$).
To suppress spontaneous emission and reduce the atomic excitation into the states $|3\rangle$ and $|4\rangle$, the one- and two-photon detunings $\Delta
_{3}=\omega _{a}-(\omega _{3}-\omega _{1})=\omega _{b}-(\omega _{3}-\omega
_{2})$ and $\Delta _{4}=\omega_{b}+\omega _{c}-(\omega _{4}-\omega _{2})$
are assumed to be large. The Zeeman sub-level $|5S_{1/2},F=1,m_{F}=1\rangle $ is far detuned from the other levels and hence can be ignored.

Based on the Hamiltonian of the system, we can derive the dimensionless nonlocal Gross-Pitaevskii equations (NGPEs)
\begin{subequations}
\label{GP1}
\begin{eqnarray}
i\partial _{t}\psi _{\uparrow } &=&-\frac{1}{2}\nabla ^{2}\psi _{\uparrow
}+k_{L}\left( \partial _{x}-i\partial _{y}\right) \psi _{\downarrow }  \\
&&-\psi
_{\uparrow }(\mathbf{r})\int d^{3}r^{\prime }\,R(\mathbf{r}-\mathbf{r^{\prime }})[|\psi
_{\uparrow }(\mathbf{r^{\prime }})|^{2}+|\psi _{\downarrow }(\mathbf{%
r^{\prime }})|^{2}], \nonumber\\
i\partial _{t}\psi _{\downarrow } &=&-\frac{1}{2}\nabla ^{2}\psi
_{\downarrow }-k_{L}\left( \partial _{x}+i\partial _{y}\right) \psi
_{\uparrow }  \\
&&-\psi _{\downarrow }(\mathbf{r})\int d^{3}r^{\prime }\,R(\mathbf{r}-\mathbf{r^{\prime }})[|\psi
_{\uparrow }(\mathbf{r^{\prime }})|^{2}+|\psi _{\downarrow }(\mathbf{%
r^{\prime }})|^{2}], \nonumber
\end{eqnarray}
\end{subequations}
where $\mathbf{r}=(x,y,z)$, $\nabla =(\partial _{x},\partial _{y},\partial _{z})$,  $d^{3}{r}^{\prime }=dx^{\prime }dy^{\prime }dz^{\prime }$,
$\psi _{\uparrow }$ and $\psi _{\downarrow }$ are respectively the wavefunctions of the spin components corresponding to the states $|1\rangle $ and $|2\rangle$. The terms $\pm k_{L}\left( \partial _{x} \mp i\partial _{y}\right) \psi _{\downarrow,\uparrow}$  represent the 2D Rashba SOC~\cite{Galitski2008,Zhai2010,Sherman2015}, with $k_{L}$ the SOC strength. The response function defining the nonlocal nonlinearity (due to Rydberg interaction between the atoms) is given by
$R(\mathbf{r}-\mathbf{r^{\prime }})=\tilde{C}_{6}/(\rho _{c}^{6}+|\mathbf{r}-%
\mathbf{r^{\prime }}|^{6})$,
with $\tilde{C}_{6}$ the modified dispersion parameter and $\rho _{c}$
the Rydberg blockade radius.
The explicit expression of the system Hamiltonian and the derivation of NGPEs (\ref{GP1}) are given in Sec.~S2 of SM~\cite{SM}.

\textit{Scaling analysis.-} To explain the stabilization
mechanism exploited here, we first resort to a scaling analysis similar
to that developed for the GPE system with local nonlinearity~\cite{Zhang2015}.
If $L$ is the characteristic size of the BEC, an estimate for the wavefunction amplitudes subject to the normalization $\int d^{3}r(|\psi _{\uparrow }|^{2}+|\psi _{\downarrow
}|^{2})=1$ gives $\psi _{\uparrow ,\downarrow }\sim L^{-3/2}$. The total
energy of the system described by the NGPEs (\ref{GP1}) includes the kinetic, SOC,
and interaction terms, i.e. $E_{%
\mathrm{tot}}=E_{\mathrm{kin}}+E_{\mathrm{soc}}+E_{\mathrm{int}}$. Here
\begin{align}
E_{\mathrm{kin}}=&\frac{1}{2}\int d^{3}r\Psi ^{\dag }\hat{\mathbf{p}}%
^{2}\Psi ,\quad E_{\mathrm{soc}}=k_{L}\int d^{3}r\Psi ^{\dag }(\hat{\mathbf{p%
}}\cdot \bm{\sigma}) \Psi ,  \notag \\
E_{\mathrm{int}}=&\frac{1}{2}\int d^{3}r\left\{ |\psi _{\uparrow }(\mathbf{r%
})|^{2}\int d^{3}r^{\prime }R(\mathbf{r}-\mathbf{r^{\prime }})\left[|\psi
_{\uparrow }(\mathbf{r^{\prime }})|^{2}+2|\psi _{\downarrow }(\mathbf{r^{\prime }})|^{2}\right] \right.  \notag \\
& \left. +|\psi
_{\downarrow }(\mathbf{r})|^{2}\int d^{3}r^{\prime }R(\mathbf{r}-\mathbf{%
r^{\prime }})\left[|\psi _{\downarrow }(\mathbf{r^{\prime }})|^{2}+2|\psi
_{\uparrow }(\mathbf{r^{\prime }})|^{2}\right]\right\} , \notag
\end{align}%
where $\Psi =(\psi _{\uparrow },\psi _{\downarrow })^{T}$, $\hat{\mathbf{p}}%
=-i(\partial _{x},\partial _{y})$, and $\bm {\sigma}=(\sigma _{x},\sigma
_{y})$, with $\sigma _{x,y}$ Pauli matrices. Then, in the weakly
nonlocal regime ($\rho _{c}\ll 1$), we obtain the scaling relations
\begin{equation}
E_{\mathrm{tot}}\sim c_{\mathrm{kin}}L^{-2}-c_{\mathrm{soc}}k_{L}L^{-1}-c_{%
\mathrm{int,0}}L^{-3}+c_{\mathrm{int,1}}L^{-5},  \label{E1}
\end{equation}%
with $c_{\mathrm{kin}}$, $c_{\mathrm{soc}}$, $c_{\mathrm{int,0}}$ and $c_{%
\mathrm{int,1}}$ positive coefficients.
In the strongly nonlocal regime ($\rho _{c}\gg 1$), the scaling relation reads
\begin{equation}
E_{\mathrm{tot}}\sim c_{\mathrm{kin}}L^{-2}-c_{\mathrm{soc}}k_{L}L^{-1}+c_{%
\mathrm{int,2}}L^{2},  \label{E2}
\end{equation}%
where $c_{\mathrm{int,2}}$ is another positive coefficient. If $E_{%
\mathrm{tot}}$ has a local minimum at a finite $L=L_{\min }$, the system may
allow a stable self-trapped state.

From the scaling relations (\ref{E1}) and (\ref{E2}), we see that both the SOC and
the nonlocal nonlinearity contribute to the stability of the self-trapped
condensate. Shown in Fig.~\ref{fig1}(b1) [\ref{fig1}(b2)] is $E_{\mathrm{tot}%
}$ in the weakly (strongly) nonlocal regime as a function of $L$ by fixing $%
c_{\mathrm{kin}}=1$ and $c_{\mathrm{int,0}}=0.2$. In the weakly (strongly)
nonlocal limit, when $c_{\mathrm{soc}}k_{L}=c_{\mathrm{int,1}}=0$ ($c_{%
\mathrm{soc}}k_{L}=c_{\mathrm{int,2}}=0$), $E_{\mathrm{tot}}$ has no local
minimum at finite $L$; however, a minimum exists when $c_{\mathrm{soc}%
}k_{L}>0$ or $c_{\mathrm{int,1}}>0$ ($c_{\mathrm{soc}}k_{L}>0$ or $c_{%
\mathrm{int,2}}>0$).

\textit{3D high-vorticity VSs and their stability.-} We
aim to derive a variational approximation (VA) to predict stable 3D VSs, as
solutions of the NGPEs (\ref{GP1}), in an accurate quasi-analytical form.
Assuming axial symmetry of the self-trapped states (obviously, it is the
highest symmetry admitted by SOC) and using cylindrical coordinates $%
(r,\varphi ,z)$, the stationary wavefunction with chemical potential $\mu $, $\psi _{\uparrow ,\downarrow }(r,\varphi ,z,t)=e^{-i\mu
t}u_{1,2}(r,\varphi ,z)$, is approximated by the\ Gaussian ansatz
\begin{equation}
u_{j}(r,\varphi ,z)=A_{j} r^{S_j}\exp \left(i S_j\varphi
-r^{2}/w_{r,j}^{2}-z^{2}/w_{z}^{2}+i\theta _{j}\right) ,  \label{ansatz}
\end{equation}%
$j=1,2$. Here $S_{j}$, $A_{j}$,  $w_{r,j}$, and $\theta _{j}$  are respectively
the integer winding number, amplitude, transverse width, and phase shift of the $j$th spin component, with $w_{z}$ the longitudinal width.
Substituting (\ref{ansatz}) into NGPEs (\ref{GP1}) demonstrates that $S_{j}$  satisfy the \emph{exact} relation (whose validity is not predicated
by VA)
$S_{2}=S_{1}+1$,
with $S_{1}=0,\,1,\,2,\cdots $. When $S_{1}=0$, the solution is a \textit{semi-vortex}
soliton, which contains a fundamental soliton in the spin-up component and a VS with $S_{2}=1$ in the spin-down one.
While the solutions with $S_{1}=1,\,2,\,\cdots $
represent the \textit{excited states} corresponding to high-order VSs \cite%
{Sakaguchi2014a,Zhang2015,Chenzhou}. In the SOC systems with local
nonlinearity, only fundamental semi-vortex solitons are stable, in the
2D~\cite{Sakaguchi2014a} and 3D~\cite{Zhang2015} cases alike, while all
excited states are unstable (they may be stabilized in the 2D system
with opposite signs of the cubic self-interaction in the two components \cite%
{Chenzhou}, or in the one with an effective nonlocal nonlinear potential,
induced by the spatial modulation of the local strength of the isotropic
repulsive dipole-dipole interaction~\cite{dip-dip}).

The Lagrangian density of the system described by NGPEs (\ref{GP1}) is
$\mathcal{L}=\mathcal{L}_{1}+\mathcal{L}_{2}$, with $\mathcal{L}_{j}$=$\mu
|u_{j}|^{2}-(|\partial _{r}u_{j}|^{2}+|\partial _{z}u_{j}|^{2})/2\pm
k_{L}e^{\mp i\varphi }u_{j}^{\ast }[\partial _{r}\mp (i/r)\partial _{\varphi
}]u_{3-j}$+$(|u_{j}|^{2}/2)[\int d^{3}rR(\mathbf{r}-\mathbf{r}^{\prime
})(|u_{j}(\mathbf{r}^{\prime 2}+|u_{3-j}(\mathbf{r}^{\prime 2})]$. The
averaged Lagrangian can be obtained by substituting ansatz~(\ref{ansatz})
into $\mathcal{L}$ and integrating it over space, i.e. $L=\int_{-\infty
}^{+\infty }d^{3}r\mathcal{L}$. Then, the equations for the variational
parameters (i.e., $A_{j}$, $w_{r,j}$, $w_{z}$ and $\theta _{j}$) can be derived
by the corresponding Euler-Lagrange equations. For details, see Sec.~S3 of SM~\cite{SM}.

\begin{figure}[tbh]
\centering
\includegraphics[width=0.95\columnwidth]{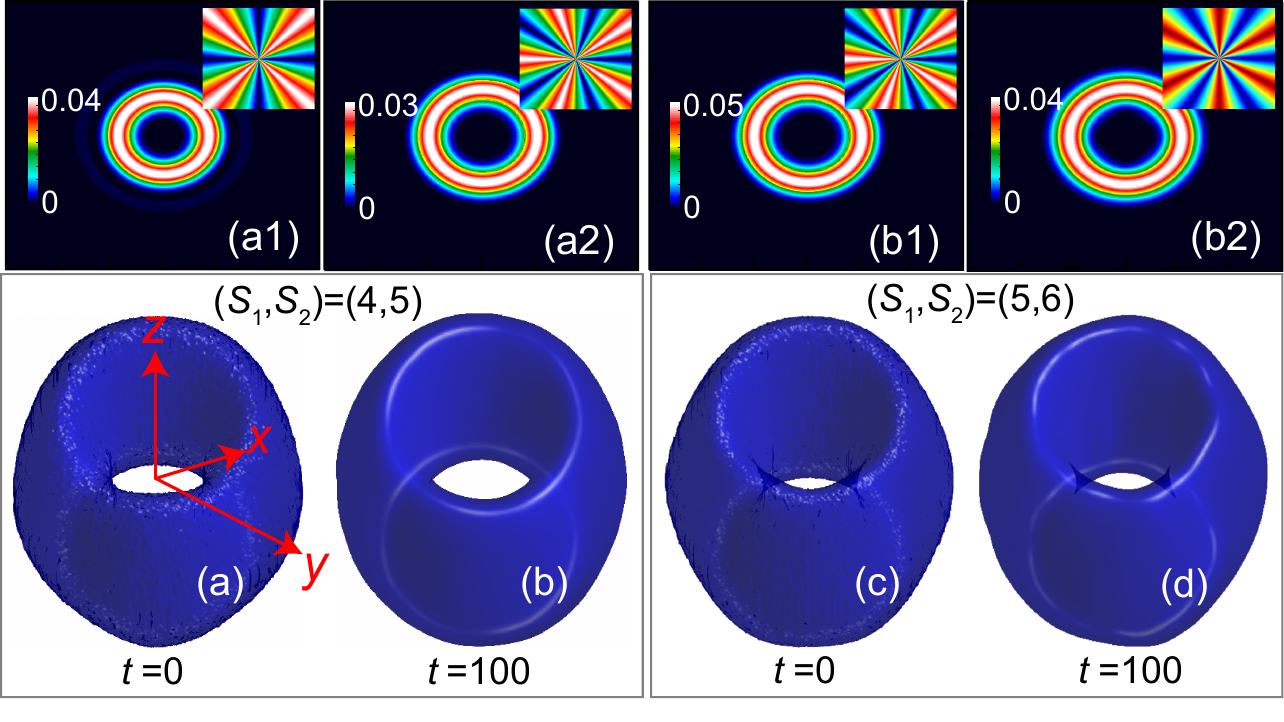} \hspace{0.5ex} \centering
\caption{Stable 3D VSs with the toroidal profiles.
(a1) and (a2) Density profiles of spin-up ($|\protect\psi _{\uparrow }|^{2}$) and spin-down ($|\protect\psi _{\downarrow }|^{2}$) components for the VS with winding
numbers $(S_{1},S_{2})=(4,5)$ as functions of $x$ and $y$.
The upper right corners give the corresponding phase profiles.
(b1) and (b2) The same as (a1) and (a2), but for $(S_{1},S_{2})=(5,6)$.
(a)~Total 3D density (toroidal) profile of the VS, $|\protect\psi _{\uparrow}|^{2}+|\protect\psi _{\downarrow }|^{2}$, at $t=0$ for $(S_{1},S_{2})=(4,5)$.
(b) The same as (a) but for $t=100$, corroborating the stability of the VS.
(c) and (d) The same as (a) and (b) but for $(S_{1},S_{2})=(5,6)$.
Parameters used are $k_{L}\, ({\rm SOC\,\,strength})=2$, $\protect\rho _{c}\,
({\rm  Rydberg\,\,blockade\,\, radius})=4$, and $\tilde{C_{6}}/\protect%
\rho _{c}^{6}\, ({\rm strength\,\, of\,\, the\,\, nonlocal\,\, nonlinearity })=1.5$.
}
\label{fig2}
\end{figure}

Shown in Fig.~\ref{fig2} are profiles of \emph{stable} 3D VSs with different topological
charges, obtained by imaginary-time propagation method~\cite{Bao} for solving NGPEs (\ref{GP1}) with the input (\ref{ansatz}).
Panels (a1) and (a2) [(b1) and (b2)] display, respectively, $|\psi _{\uparrow }|^{2}$ and $|\psi
_{\downarrow }|^{2}$ as functions of $x$ and $y$ at $z=0$ cross-section, for winding numbers $(S_{1},S_{2})=(4,5)$ [$(S_{1},S_{2})=(5,6)$]. The corresponding phase profile is given in the upper right corner of each panel.
Panel (a) gives the total 3D density profile (i.e. $|\protect\psi _{\uparrow}|^{2}+|\protect\psi _{\downarrow }|^{2}$) at $t=0$ for $(S_{1},S_{2})=(4,5)$; panel (b) is the same as (a) but for $t=100$,
the result for testing the stability of the VS by real-time simulations of Eqs.~(\ref{GP1}), obtained by taking the panel (a) as an input with the addition of a random perturbation at  $5\%$ level.
Plotted in panels (c) and (d)  are same as panels (a) and (b) but for $(S_{1},S_{2})=(5,6)$.  The sets of the 3D profiles shown here corroborates clearly the stability of the high-order VSs. In order to know the relation between the topological charge and total energy of VSs, we calculate $E_{\rm tot}$ with different $S_1$, which increases monotonously with $S_{1}$. Thus, the semi-vortex soliton has the lowest energy while high-order VSs carry larger energies (for more detail, see Sec.~S4 of SM~\cite{SM}).

\begin{figure}[tbh]
\centering
\includegraphics[width=0.95\columnwidth]{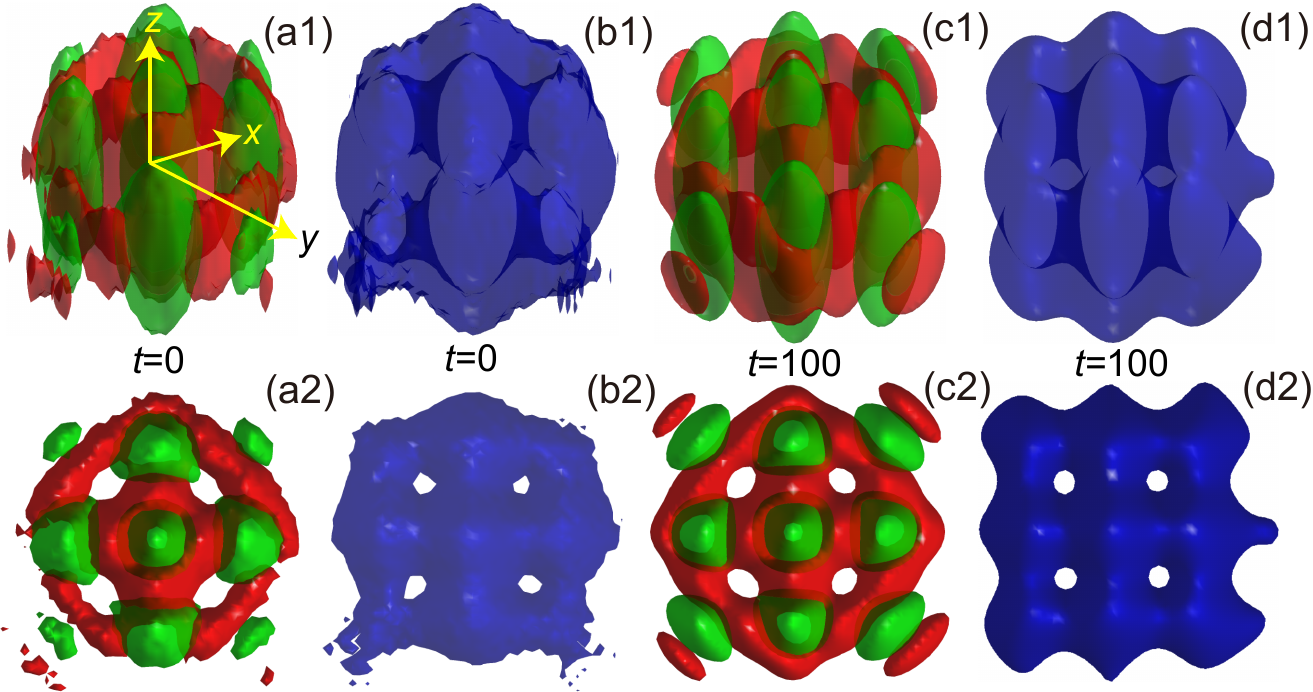}
\caption{Stable 3D VS with the chessboard-like profile for $(S_{1},S_{2})=(4,5$).
(a1)~3D density profiles of the VS for the spin-up ($|\protect\psi
_{\uparrow }|^{2}$; red color) and spin-down ($|\protect\psi _{\downarrow
}|^{2}$; green color) components at $t=0$.
(b1)~The same as (a1) but for the total-density
profile of the VS (i.e. $|\protect\psi _{\uparrow }|^{2}+|\protect\psi _{\downarrow }|^{2}$).
(c1) and (d1) Evolution results of the VS [corresponding to (a1) and (b1)] at $t=100$
in the presence of perturbations. The respective top views of (a1)-(d1)
are displayed in panels (a2)-(d2). Parameters used are the same as in Fig.~(\protect\ref{fig2}) except $k_{L}=3.5$. }
\label{fig3}
\end{figure}

When the SOC strength $k_{L}$ exceeds a critical value, \textit{viz}., $%
k_{L}\gtrsim 2.6$, we find that the VS shape transforms from the toroidal profile into a
chessboard-like one, due to the spontaneous breaking of the rotational symmetry
in the transverse ($x,y$) plane. At the same time, such a VS with the broken rotational symmetry exhibits separation of its components, as their densities occupy different spatial domains, with nearly no overlap between them (see  Sec.~S4 of SM~\cite{SM} for more detail). These features are illustrated in Fig.~\ref{fig3}(a1) by the
density profiles of the spin-up ($|\psi _{\uparrow }|^{2}$) and spin-down ($%
|\psi _{\downarrow }|^{2}$) components at $t=0$ for $(S_{1},S_{2})=(4,5)$ and $k_{L}=3.5$.
In Fig.~\ref{fig3}(b1), the total density profile ($|\psi
_{\uparrow }|^{2}+|\psi _{\downarrow }|^{2}$) is plotted for the same VS. The stability of the chessboard-like VS is corroborated
by the result of its perturbed evolution at $t=100$, displayed in
Figs.~\ref{fig3}(a1)-(d1). The shape and stability of the same VS is
additionally illustrated by their top views in Figs.~\ref{fig3}(a2)-(d2). Note that the chessboard-like VS patterns displayed here do not exhibit any rotation.
From the results of Figs.~\ref{fig2} and \ref{fig3} we see that the interplay of SOC and the nonlocal Rydberg nonlinearity secures the full stability of the 3D VSs with high
topological charges in free space (i.e. without external potential), including the immunity of the solitons to the azimuthal instability (which is usually most difficult to provide~\cite{Malomed2019}).

To check the stability of the VSs further, a systematic investigation on the linear stability analysis of the VSs is performed by  taking  $\psi _{\uparrow ,\downarrow }$=$e^{-i\mu t+iS_{1,2}\varphi }[\phi _{1,2}(r,z)+p_{1,2}(r,z)e^{\lambda
t}e^{i\kappa \varphi }+q_{1,2}^{\ast }(r,z)e^{\lambda ^{\ast }t}e^{-i\kappa
\varphi }]$, where $\phi _{j}(r,z)$ ($j=1,2$) are the numerically found stationary
profiles, $p_{j}(r,z)$ and $q_{j}(r,z)$ represent
eigenmodes of small perturbation, $\lambda $ is the
perturbation growth rate, and integer $\kappa $ is the azimuthal
perturbation index. By substituting the ansatz into the NGPEs (\ref{GP1})
and linearizing the equations with respect to $p_{j}$ and $%
q_{j}$, we arrive at an eigenvalue problem that can be solved numerically; see Sec.~S5 of SM~\cite{SM} for detail. The VS is stable if all
eigenvalues are purely imaginary [i.e. Re$(\lambda )= 0$], unstable otherwise.

\begin{figure}[tbh]
\centering
\includegraphics[width=1.0\columnwidth]{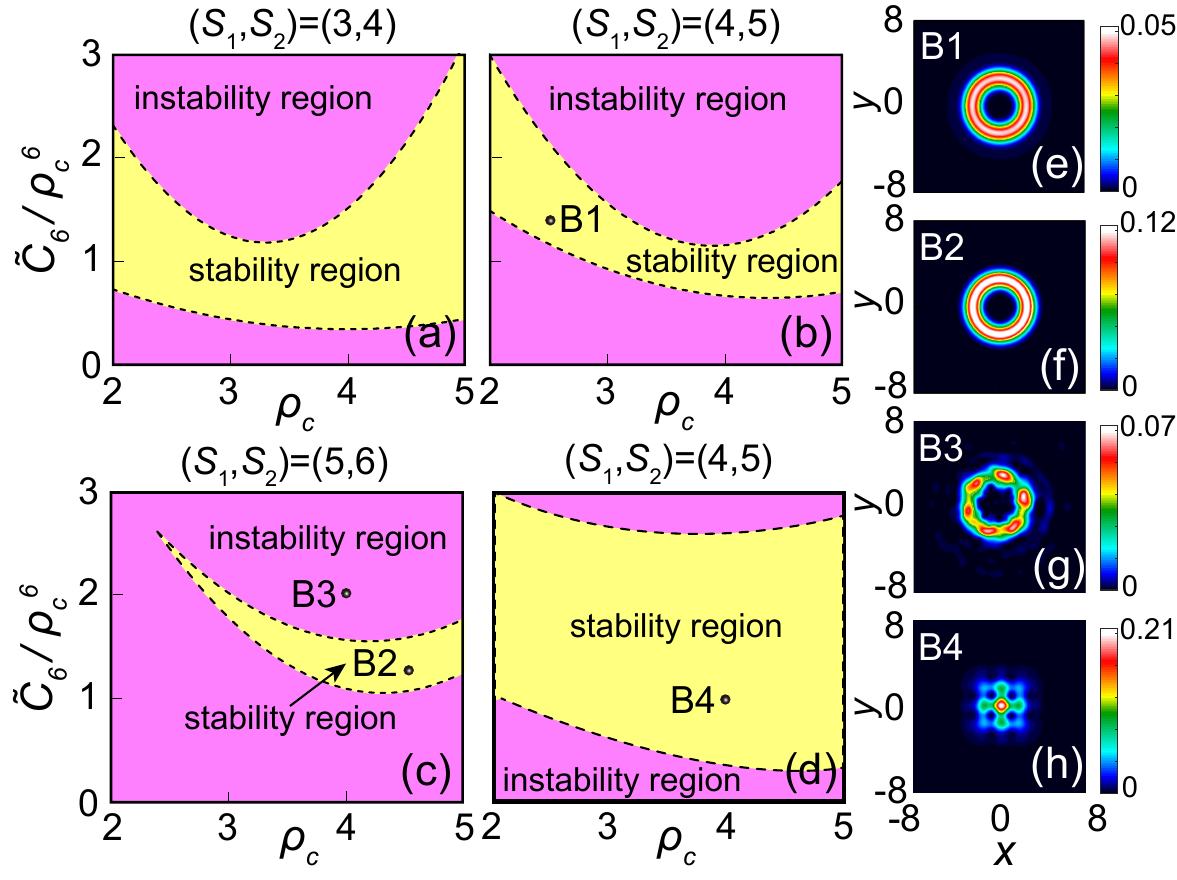}
\caption{Stability charts and shapes of stable and unstable
3D VSs. (a), (b), and (c) Stability (yellow) and instability (red) domains of the toroidal VSs in the plane of $\protect\rho _{c}$ and $\tilde{C_{6}}/\protect\rho _{c}^{6}$, for $(S_{1},S_{2})=(3,4)$, $(4,5)$ and $(5,6)$, respectively. (d) Stability and instability domains of the chessboard-like VS. Points B1, B2, B3 and B4 in panels (b), (c) and (d) correspond to $(\protect\rho _{c},\tilde{C_{6}}/\protect\rho _{c}^{6})=(2.5,1.5)$, $(4.5,1.3)$, $(4,2.1)$ and $(4,1)$, respectively. (e), (f), (g) and (h)  Total-density profiles $|\protect\psi_{\uparrow }|^{2}+|\protect\psi _{\downarrow }|^{2}$ of stable [panels (e), (f) and (h)] and unstable [panel (g)] VSs in the cross section $z=0$,
corresponding to the points B1, B2, B3, and B4 in panels
(b), (c) and (d), respectively. Here, the stable (unstable) VSs are obtained at $t=100$ ($t=30$). The SOC strength  is taken to be $k_{L}=2$ for panels (a)-(c), and $k_{L}=3.5$ for panel (d).
}
\label{fig4}
\end{figure}

Shown in Figs.~\ref{fig4}(a), \ref{fig4}(b), and \ref{fig4}(c)
are stability charts of the toroidal VSs, respectively for $(S_{1},S_{2})=(3,4)$, $(4,5)$, and $(5,6)$ in the plane of $\rho _{c}$ and $\tilde{C_{6}}/\rho _{c}^{6}$. We see that, for all values of $\rho _{c}$, the VSs are
stable for moderate values of $\tilde{C_{6}}/\rho _{c}^{6}$ (yellow
domains); when $\tilde{C_{6}}/\rho _{c}^{6}$ is small (the
Rydberg interaction is too weak), the VSs are unstable, spreading out during
the evolution. On the other hand, when $\tilde{C_{6}}/\rho _{c}^{6}$ is
large (the Rydberg interaction is too strong), the azimuthal instability breaks the
unstable VSs into sets of fragments, which eventually blow up through
intrinsic collapse (not shown here). The stability domain shrinks when the winding numbers $S_{1,2}$ increase and almost vanishes when $S_{1}\geq6$ ($S_{2}\geq7$).

Shown in Fig.~\ref{fig4}(d) is the stability chart for the chessboard-like VS with $(S_{1},S_{2})=(4,5)$. It is seen that the stability domain is much larger than for the toroidal VSs, as the chessboard-shaped states are immune to the azimuthal instability. Moreover, the winding numbers have a marginal effect on the stability domain (therefore the results for other winding numbers are skipped). Panels (e), (f), (g) and (h) in Fig.~\ref{fig4} display density profiles of stable [panels (e), (f) and (h)] and unstable [panel (g)] VSs, corresponding to points $B_{1,2,3,4}$, in Figs.~\ref{fig4}(b), \ref{fig4}(c) and \ref{fig4}(d), respectively.

\begin{figure}[tbh]
\centering
\includegraphics[width=1.0\columnwidth]{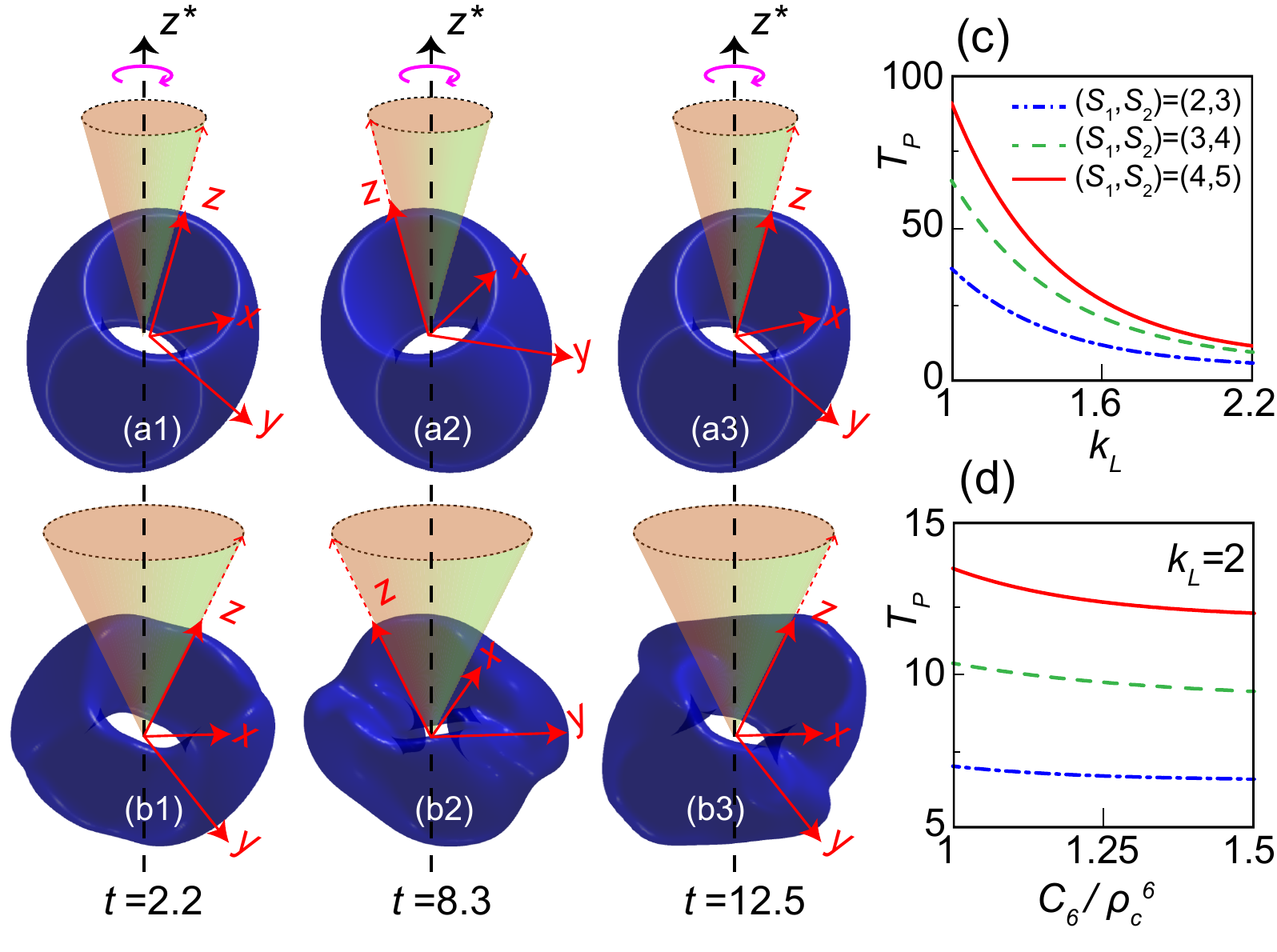}
\caption{VS gyroscopes. (a1), (a2), and (a3) Isosurface plots of the total density profile $|\protect\psi _{\uparrow }|^{2}+|\protect\psi _{\downarrow }|^{2}$ for $%
(S_{1},S_{2})=(3,4)$ at $t=2.2$, $8.3$ and $12.5$, respectively.
The VS precession is initiated by a relatively weak torque.
(b1), (b2), and (b3) The same as (a1), (a2), and (a3) but for a relatively strong torque. (c)~The precession period $T_{p}$ as a function of $k_{L}$ for $(\protect%
\rho _{c},\tilde{C_{6}}/\protect\rho _{c}^{6})=(4,1.5)$. (d)~$T_{p}$ as a function
of $\tilde{C_{6}}/\protect\rho _{c}^{6}$ for $k_{L}=2$. The red solid, green
dashed, and blue dot-dashed lines correspond to $(S_{1},S_{2})=(2,3)$, $%
(3,4) $ and $(4,5)$, respectively.}
\label{fig5}
\end{figure}

\textit{VS gyroscopes and related applications.-} The stable 3D
VSs of toroidal shape obtained above feature robust dynamics similar to
that of mechanical gyroscopes. To demonstrate this, we apply a torque to the
VS, whose axle is along the $z$ direction, multiplying it by factor
$T=\exp \left[ i\alpha z\tanh (x/x_{0})\right]$,
with  $\alpha $ the strength and $x_{0}$ the transverse size~\cite
{Driben2014a}. A relatively weak torque ($\protect\alpha =0.1$, $x_{0}=10$) gives rises to periodic precession of the axle of the VS torus [see Fig.~\ref{fig5}(a1)-(a3)]. However, a strong torque ($\protect\alpha =0.5$, $x_{0}=10$) deforms the VS, although it keeps the vorticity, along with the inner hole [Fig.~\ref{fig5}(b1)-(b3)]. In the course of the
subsequent evolution, the deformed toroidal VS gradually restores its shape.

The precession period $T_{p}$ of the torque-kicked VS depends on winding
numbers, SOC strength, and Rydberg interaction strength $\tilde{C_{6}}/\rho _{c}^{6}$. Plotted in panels (c) and (d) of Fig.~\ref{fig5} is $T_{p}$ for $(S_{1},S_{2})=(2,3)$, $(3,4)$, and $(4,5)$, respectively. We see that $T_{p}$
decreases monotonously as $k_{L}$ and $\tilde{C_{6}}/\rho _{c}^{6}$
increase, or the winding numbers decrease. Thus, the precession of the VS
gyroscopes may be exploited to measure the dispersion parameter $C_{6}$ and
blockade radius $\rho _{c}$ of the Rydberg state.

\textit{Conclusion.-} We have investigated the existence and
stabilization of high-order 3D VSs in the bimodal
Rydberg-dressed BEC in the presence of SOC. We have shown that the stable VSs exist in free space in the broad area of the system's parameter domains with the winding numbers up to $(S_1, S_2) =( 5, 6)$. When the SOC strength exceeds the critical value, the rotational symmetry of the VSs is broken and their components tend to separate, but without destabilizing them. We have also shown that the application
of the torques to the VSs sets them in the state of gyroscopic precession.

{\it Acknowledgments.-}
This work is supported by NSF of China
under Grant No.~12374303, and in part by the Israel Science Foundation through grant No.~1695/22.


\end{document}